\documentclass[11pt,reqno,oneside]{article}
\usepackage{pstricks,pst-node,pst-tree}
\usepackage{amsmath}
\usepackage{amssymb}
\usepackage{color}
\usepackage[lmargin=.55in, rmargin=.55in, tmargin=1in, bmargin=1in]{geometry}
\usepackage{xcolor}
\usepackage[dvips]{graphicx}

\usepackage{natbib}

\usepackage{pstricks}

\newtheorem{prop}{Proposition}
\newtheorem{coro}{Corollary}

\definecolor{myblue}{rgb}{0,0,0}
\definecolor{myred}{rgb}{0.9,0.1,0}
\newcommand{\fil}[1]{\textcolor{myblue}{#1}}

\title{Enumeration of ancestral configurations \\ for matching gene trees and species trees}
\author{Filippo Disanto\thanks{Corresponding author. Email: fdisanto@stanford.edu.} $\,$ and Noah A.~Rosenberg \\
\\ {\small Department of Biology, Stanford University, Stanford, CA 94305 USA} }

\begin{document}

\maketitle

\begin{abstract}
Given a gene tree and a species tree, \emph{ancestral configurations} represent the combinatorially distinct sets of gene lineages that can reach a given node of the species tree. They have been introduced as a data structure for use in the recursive computation of the conditional probability under the multispecies coalescent model of a gene tree topology given a species tree, the cost of this computation being affected by the number of ancestral configurations of the gene tree in the species tree. For matching gene trees and species trees, we obtain enumerative results on ancestral configurations. We study ancestral configurations in balanced and unbalanced families of trees determined by a given seed tree, showing that for seed trees with more than one taxon, the number of ancestral configurations increases for both families exponentially in the number of taxa~$n$. For fixed $n$, the maximal number of ancestral configurations tabulated at the species tree root node and the largest number of labeled histories possible for a labeled topology occur for trees with precisely the same unlabeled shape. For ancestral configurations at the root, the maximum increases with $k_0^n$, where $k_0 \approx 1.5028$ is a quadratic recurrence constant. Under a uniform distribution over the set of labeled trees of given size, the mean number of root ancestral configurations grows with $\sqrt{3/2}(4/3)^n$ and the variance with approximately $1.4048(1.8215)^n$. The results provide a contribution to the combinatorial study of gene trees and species trees.

\end{abstract}


\section{Introduction}


Investigations of the evolution of genomic regions along species tree branches have generated new combinatorial structures that can assist in studying gene trees and species trees \citep{Maddison97, DegnanAndSalter05, ThanAndNakhleh09, DegnanEtAl12:mathbiosci, Wu12}. Among these structures are \emph{ancestral configurations}, structures that for a given gene tree topology and species tree topology represent the possible sets of gene lineages that can reach a given node of the species tree \citep{Wu12}.

Ancestral configurations represent the set of objects over which recursive computations are performed in a fundamental calculation for inference of species trees from information on multiple genetic loci: the evaluation of gene tree probabilities conditional on species trees \citep{Wu12}. Because of the appearance of ancestral configurations in sets over which sums are computed (e.g.~eq.~(7) of \cite{Wu12}), solutions to enumerative problems involving ancestral configurations contribute to an understanding of the computational complexity of phylogenetic calculations.

Under the assumption that a gene tree and a species tree have a matching labeled topology $t$, we examine the number of ancestral configurations that can  appear at the nodes of the species tree. Extending results of \cite{Wu12}\fil{, whose appendix reported the number of ancestral configurations for caterpillar species trees and established a lower bound for completely balanced species trees}, we study the number of ancestral configurations when $t$ belongs to families of trees characterized by a balanced or unbalanced pattern and a seed tree. As a special case, we derive upper and lower bounds on the number of ancestral configurations possessed by matching gene trees and species trees of given size. Finally, we study the mean and the variance of the number of ancestral configurations when $t$ is a random labeled tree of given size selected under a uniform distribution.


\section{Preliminaries}

We study ancestral configurations for rooted binary labeled trees. We start with some definitions and preliminary results. In Section~\ref{filo}, we recall basic properties of rooted binary labeled trees. In Section~\ref{acomb}, we recall properties of generating functions that will be used to derive some of our enumerative results. Following \cite{Wu12}, in Section~\ref{igno} we define ancestral configurations for matching gene trees and species trees, and we determine a recursive procedure to compute their number at a given node of a tree. We then relate the total number of ancestral configurations in a tree to the number of ancestral configurations at the root of the tree.


\subsection{Labeled topologies}
\label{filo}

A labeled topology, or ``tree'' for short, of size $n$ is a bifurcating rooted tree with $n$ labeled taxa (Fig.~\ref{config}A). We assume without loss of generality a linear (alphabetical) order $a \prec b \prec c \prec  ... $ among the set $\{a,b,c,...\}$ of possible labels for the taxa of a tree. A tree of size $n$ has leaves labeled using the first $n$ labels in the order $\prec$.
Given two trees $t_1$ and $t_2$, we write $t_1 \cong t_2$ and say that $t_1$ is isomorphic to $t_2$ when, removing labels at their taxa, $t_1$ and $t_2$ share the same unlabeled topology. The set of trees of size $n$ is denoted by $T_n$, and $T = \bigcup_{n\geq 1} T_n$ denotes the set of all trees of any size. The number of trees of size $n \geq 2$ can be computed as $|T_n|=(2n-3)!! = 1 \times 3 \times 5 \times ... \times (2n-3)$
\citep{Felsenstein78},
which can be rewritten for $n \geq 1$ as
\begin{equation}\label{carciofo}
|T_n| =
\frac{(2n-2)!}{2^{n-1} (n-1)!} =
\frac{(2n)!}{2^n(2n-1)  n!}.
\end{equation}
The exponential generating function associated with the sequence $|T_n|$ is defined as
\begin{equation}\label{genio}
T(z)= \sum_{t \in T} \frac{z^{|t|}}{|t|!} = \sum_{n=1}^\infty \frac{|T_n| z^n}{n!} = z + \frac{z^2}{2} + \frac{3z^3}{6} + \frac{15z^4}{24} + \ldots , \end{equation}
and it is given by \cite{FlajoletAndSedgewick09} (Example II.19)
\begin{equation}\label{expo}
T(z) = 1-\sqrt{1-2z}.
\end{equation}


Throughout the paper, although most of our results are purely combinatorial, where a probability distribution on the set of labeled topologies of a given size is needed, we assume a uniform probability distribution over the set of trees of given size.



\subsection{Exponential growth and analytic combinatorics}
\label{acomb}

Following \cite{FlajoletAndSedgewick09},
a sequence of non-negative numbers $a_n$ is said to have exponential growth $k^n$ or, equivalently, to be of exponential order $k$, when
\begin{equation*}\label{limosup}
\limsup_{n \rightarrow \infty} \big[(a_n)^{1/n}\big] = \lim_{n \rightarrow \infty} \big[\sup_{m \geq n} \big[(a_m)^{1/m}\big] \big] = k.
\end{equation*}
This relation can be rephrased as $a_n = k^n s(n)$, where $s$ is a subexponential factor, that is, $\limsup_{n \rightarrow \infty} [s(n)^{1/n}] = 1$. By these definitions, a sequence $a_n$ grows exponentially in $n$ when its exponential order strictly exceeds 1.

The exponential order of a sequence gives basic information about its speed of growth and enables comparisons with other sequences. In particular, from the definition, it follows that if $(a_n)$ has exponential order $k_a$ and $(b_n)$ has exponential order $k_b < k_a$, then the sequence of ratios $b_n/a_n$ converges to $0$ exponentially fast as $(k_b/k_a)^n$. If two sequences $(a_n)$ and $(b_n)$ have the same exponential growth, then we write $a_n \bowtie b_n$.

We are interested in the exponential growth of several increasing sequences of non-negative integers. Several results will be obtained through techniques of analytic combinatorics (see Sections IV and VI of \cite{FlajoletAndSedgewick09}). The entries of a sequence of integers $(a_n)_{n \geq 0}$ can be interpreted as the coefficients of the power series expansion $A(z) = \sum_{n=0}^{\infty} a_n z^n$ at $z=0$ of a function $A(z)$, the generating function of the sequence. Considering $z$ as a complex variable, under suitable conditions,
there exists a general correspondence between the singular expansion of the generating function $A(z)$ near its dominant singularity---the one nearest to the origin---and the asymptotic behavior of the associated coefficients $a_n$. In particular, the exponential order of the sequence $(a_n)$ is given by the inverse of the modulus of the dominant singularity of $A(z)$. For instance, the exponential order of the sequence $|T_n|/n!$, where $|T_n|$ is as in (\ref{carciofo}), is given by $2$ because $1/2$ is the dominant singularity of the associated generating function (\ref{expo}). In other words, $|T_n|/n!$ increases with a subexponential multiple of $2^n$ as $n$ becomes large.


\subsection{Gene trees, species trees, and ancestral configurations}
\label{igno}

In this section, we define the object on which our study focuses: the ancestral configurations of a gene tree $G$ in a species tree $S$. Ancestral configurations have been introduced by \cite{Wu12}. In our framework, where exactly one gene lineage has been selected from each species, we assume $G$ and $S$ to have the same labeled topology $t$.


\subsubsection{Ancestral configurations}
\begin{figure}
\begin{center}
\includegraphics*[width=0.86\textwidth,trim=0 0 0 0]{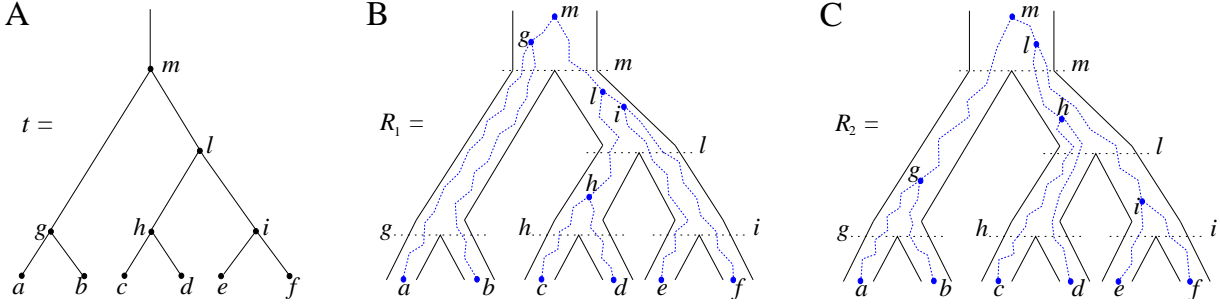}
\end{center}
\vspace{-.7cm}
\caption{{\small  A gene tree and a species tree with a matching labeled topology $t$. {\bf(A)} A tree $t$ of size $6$ isomorphic to the gene tree and species tree depicted in (B) and (C). Tree $t$ is characterized by its shape and by the labeling of its taxa. It is convenient to label the internal nodes of $t$, by $g,h,i,\ell,m$ in this case. We identify each lineage (edge) of $t$ by the lowest node it intersects, so lineage $g$ results from the coalescence of lineages $a$ and $b$.
{\bf (B)} A possible realization $R_1$ of the gene tree in (A) (dotted lines) in the species tree with a matching topology (solid lines). The ancestral configuration at species tree node $\ell$ is $\{ h,e,f \}$. The configuration at node $m$ is $\{ a,b,\ell \}$.  {\bf (C)} A different realization $R_2$ of the gene tree in (A) in the matching species tree. The configurations at species tree nodes $\ell$ and $m$ are $\{ c,d,i \}$ and $\{ g,h,i \}$, respectively.
}} \label{config}
\end{figure}

Suppose $R$ is a realization of a gene tree $G$ in a species tree $S$, where $G=S=t$ (Fig.~\ref{config}). In other words, $R$ is one of the evolutionary possibilities for the gene tree $G$ on the matching species tree $S$. Viewed backward in time, for a given node $k$ of $t$, consider the set $C(k,R)$ of gene lineages (edges of $G$) that are present in $S$ at the point right before node $k$.

As in \cite{Wu12}, the set $C(k,R)$ is called the \emph{ancestral configuration} of the gene tree at node $k$ of the species tree. Taking the tree $t$ depicted in Fig.~\ref{config}A and considering the realization $R_1$ of the gene tree $G=t$ in the species tree $S=t$ as given in Fig.~\ref{config}B, we see that the gene lineages $a$, $b$, and $\ell$ are those present in the species tree at the point right before the root node $m$. The set $C(m,R_1) = \{ a,b,\ell \}$ is thus the ancestral configuration of the gene tree at node $m$ of the species tree. Similarly, the ancestral configuration of the gene tree at node $\ell$ of the species tree is the set of gene lineages $C(\ell,R_1)=\{ h,e,f  \}$. In Fig.~\ref{config}C, where a different realization $R_2$ of the same gene tree is depicted, the ancestral configuration at the root $m$ of the species tree is the set of gene lineages $C(m,R_2)=\{g,h,i \}$. The ancestral configuration at node $\ell$ is $C(\ell,R_2) = \{c,d,i \}$.

Let $\Re(G,S)$ be the set of possible realizations of the gene tree $G=t$ in the species tree $S=t$. For a given node $k$ of $t$, by considering all possible elements $R \in \Re(G,S)$, we define the set
\begin{equation}\label{ciccio}
C(k) = \{ C(k,R) : R \in \Re(G,S)   \}
\end{equation}
 and the number
\begin{equation}\label{cicci}
c(k) = |C(k)|.
\end{equation}
Thus, $c(k)$ corresponds to the number of different ways the gene lineages of $G$ can reach the point right before node $k$ in $S$, when all possible realizations of the gene tree $G$ in the species tree $S$ are considered. For instance, taking $t$ as in Fig.~\ref{config}A we have $C(g) = \{\{ a,b \}\}, C(\ell)=\{ \{c,d,e,f  \}, \{h,e,f  \}, \{c,d,i  \},\{h,i  \}  \}$, and
\begin{equation}\label{cm}
\small{
C(m) = \{ \{g,\ell \}, \{a,b,\ell  \}, \{g,c,d,e,f  \},\{a,b,c,d,e,f  \}, \{g,h,e,f  \}, \{a,b,h,e,f  \}, \{g,c,d,i  \}, \{a,b,c,d,i  \} , \{g,h,i  \}, \{a,b,h,i  \} \}.
}
\end{equation}
Note that for two different realizations $R_1,R_2 \in \Re(G,S)$ \fil{and an internal node $k$}, we do not necessarily have $C(k,R_1) \neq C(k,R_2)$.

For each internal node $k$, our definition of ancestral configuration specifically excludes as a possibility the case in which all gene tree lineages descended from node $k$ have coalesced at species tree node $k$, so that $\{k \} \notin C(k)$. Each configuration at node $k$ is considered at the point right before node $k$ in the species tree, and there is thus no time for the gene lineages from the left subtree of $k$ to coalesce with those from the right subtree of $k$. Our definition is identical to that of \cite{Wu12}, with the exception that we say that a leaf or 1-taxon tree has 0 ancestral configurations, whereas Wu assigns these cases 1 ancestral configuration.

Because we assume gene tree $G$ and species tree $S$ have the same labeled topology $t$, the set $C(k)$ and the quantity $c(k)$ defined in (\ref{ciccio}) and (\ref{cicci}) depend only on node $k$ and tree $t$. In what follows, we use the term \emph{configuration at node $k$ of $t$} to denote an element of $C(k)$.
The next result provides a recursive procedure for calculating the number $c(k)$ at a given node $k$ of $t$.


\begin{prop}\label{prop1}
Given a tree $t$ with $|t|>1$, the number $c(r)$ of possible configurations at the root $r$ of $t$ can be recursively computed as
\begin{equation}\label{eqC}
c(r) = 1 + c(r_{\ell}) + c(r_r) + c(r_{\ell}) c(r_r)= [c(r_{\ell})+1][c(r_r)+1],
\end{equation}
where $r_{\ell}$ (resp.~$r_r$) denotes the left (resp.~right) child of $r$, and $c(r)$ is set to $0$ when $|t|=1$.
\end{prop}
\emph{Proof.} If $A$ and $B$ are two sets of sets, we define $A \otimes B = \{ a \cup b : a \in A, b \in B   \}$. The set $C(r)$ of configurations at internal node $r$ can be decomposed as
\begin{equation}\label{momo}
C(r) = \{ \{ r_{\ell},r_r \} \} \cup \big[ C(r_{\ell}) \otimes \{ \{r_r \}  \} \big] \cup \big[ \{ \{ r_{\ell}\} \} \otimes C(r_r) \big] \cup   \big[ C(r_{\ell}) \otimes C(r_r) \big],
\end{equation}
where the set unions are disjoint because, as already noted, $\{r_{\ell} \} \notin C(r_{\ell})$  and $\{r_{r} \} \notin C(r_{r})$. We immediately obtain (\ref{eqC}), as $c(r)=|C(r)|$. $\Box$

\bigskip

\fil{We reiterate that in order for (\ref{eqC}) to apply for all $t$ with $|t|>1$, we must set to $0$ the number of configurations at a species tree leaf and at the root of the 1-taxon tree.} For the tree depicted in Fig.~\ref{config}A, each configuration in $C(m)$ (\ref{cm}) can be obtained as described in (\ref{momo}) from the configurations in $C(g)$ and $C(\ell)$. Note indeed that $c(m)=10=(1+1)(4+1)=[c(g)+1][c(\ell)+1]$, as determined by (\ref{eqC}).

\begin{figure}
\begin{center}
\includegraphics*[scale=0.5,trim=0 0 0 0]{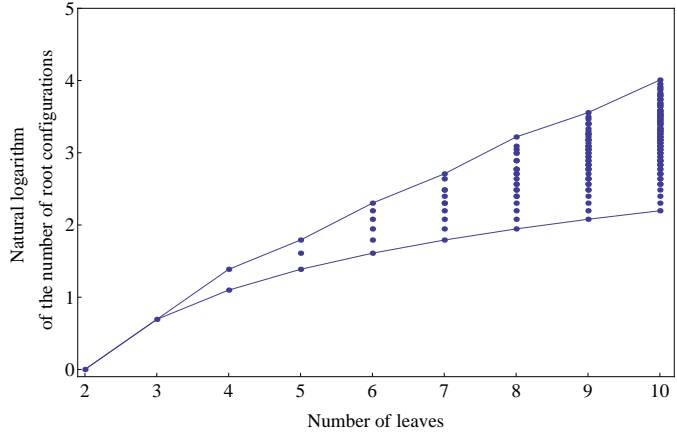}
\end{center}
\vspace{-.7cm}
\caption{{\small Natural logarithm of the number of root configurations for all possible tree shapes of size $2\leq n \leq 10$. The value for $n=1$, $\log(0)$, is omitted. Dots corresponding to the largest and smallest number of root configurations for each $n$ are connected by the top and bottom lines, respectively.
}} \label{smalltrees}
\end{figure}


\subsubsection{Total configurations and root configurations}
\label{pippi}

Let $K(t)$ be the set of nodes of a tree $t$. The number of nodes $|K(t)|$ satisfies $|K(t)| = 2|t|-1 < 2|t|$. Define the total number of configurations in $t$ as the sum $$c = \sum_{k \in K(t)} c(k).$$ Let $c(r)$ be the number of configurations at the root $r$ of $t$, or \emph{root configurations} for short. As is shown in Appendix 1, $c(r)$ satisfies the bound
\begin{equation}\label{miaomiao}
c(r) \leq 2^{|t|}/2.
\end{equation}
Furthermore, because $c(r) \geq c(k)$ for each node $k$ of $t$, we have
\begin{equation}\label{bau}
c(r) \leq c \leq 2|t| c(r).
\end{equation}
This result indicates that the total number of configurations $c$ and the number of root configurations $c(r)$ are equal up to a factor that is at most polynomial in the tree size $|t|$. A consequence is that in measuring $c(r)$ for a family $(t_i)$ of trees of increasing size, an exponential growth of the form $c(r) \bowtie k^{|t|}$ for the number of root configurations translates into the same exponential growth for the total number of configurations in $t$:
\begin{equation}\label{doct}
c(r) \bowtie k^{|t|} \Leftrightarrow c \bowtie k^{|t|},
\end{equation}
where by virtue of (\ref{miaomiao}), $ k \leq 2$.

An equivalent result holds when we consider the expected value of the total number of configurations $\mathbb{E}_n[c]$ in a random labeled tree topology of given size $n$. Indeed, when a tree of size $n$ is selected at random from the set of labeled topologies, inequality (\ref{bau}) gives $\mathbb{E}_n[c(r)] \leq \mathbb{E}_n[c] \leq 2n\mathbb{E}_n[c(r)].$ Thus, the exponential growth of $\mathbb{E}_n[c]$ with respect to $n$ can be recovered from the exponential growth of $\mathbb{E}_n[c(r)]$,
\begin{equation}\label{bau2}
\mathbb{E}_n[c] \bowtie \mathbb{E}_n[c(r)].
\end{equation}
Similarly, for the second moment $\mathbb{E}_n[c^2]$ we have $\mathbb{E}_n[c(r)^2] \leq \mathbb{E}_n[c^2] \leq 4n^2\mathbb{E}_n[c(r)^2]$, and thus
\begin{equation}\label{bau3}
\mathbb{E}_n[c^2] \bowtie \mathbb{E}_n[c(r)^2].
\end{equation}

Using these results, in Sections~\ref{families} and \ref{trandom} we will determine the exponential growth of $c(r)$ and $c$ with respect to size $|t|$, when $t$ is considered in different settings. In Section~\ref{families}, $t$ belongs to families of unbalanced or balanced trees, whereas in Section~\ref{trandom}, we perform our analysis considering $t$ as a random labeled topology of given size.

\subsection{Root configurations in small trees}

For small values of $n$, formula (\ref{eqC}) enables the exhaustive computation of the number of root configurations $c(r)$ for representative labelings of each of the unlabeled topologies of size $n$. In Fig.~\ref{smalltrees}, each dot corresponds to the logarithm of the number of root configurations for a certain tree shape of size determined by its $x$-coordinate. The dots associated with the largest values of $c(r)$ are connected by the top line, whose growth is linear in $n$. Indeed, as was shown by \cite{Wu12}, there exist families of trees for which the growth of the number of root configurations is exponential in the tree size. From (\ref{miaomiao}), it follows that the growth of the sequence of the largest number of \fil{root} configurations in trees of size $n$ must be  exponential in $n$ as well.

\begin{figure}
\begin{center}
\includegraphics*[scale=0.9,trim=0 0 0 0]{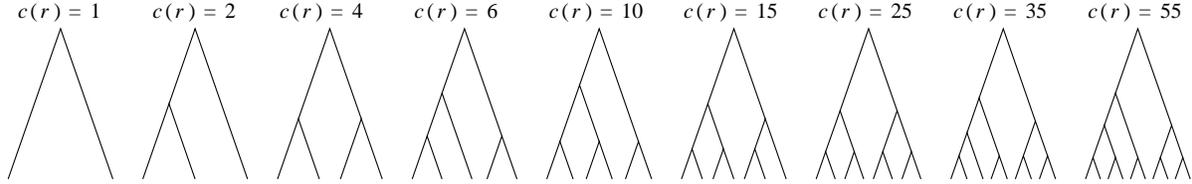}
\end{center}
\vspace{-.7cm}
\caption{{\small Tree shapes of size $2 \leq n \leq 10$ whose labeled topologies have the largest number of root configurations among trees of size $n$. The number of root configurations $c(r)$ is indicated for each tree. In each tree displayed, the two root subtrees each maximize the number of root configurations among trees of their size.
}} \label{maxi}
\end{figure}

The tree shapes whose labeled topologies possess the largest number of root configurations among trees of fixed size appear in Fig.~\ref{maxi} together with their number of root configurations $c(r)$. Starting with $n=4$, each shape in the sequence can be seen to be produced by connecting two smaller shapes also in the sequence (possibly the same shape) to a shared root.

The tree shape that minimizes the number of root configurations is the caterpillar topology. The number of root configurations in the caterpillar of size $n$ is $n-1$ \citep{Wu12}. The bottom line in Fig.~\ref{smalltrees}, which connects dots corresponding to the smallest number of root configurations for a tree with $n$ taxa, grows with $\log(n-1)$.

These observations show that tree topology can have a considerable impact on the number of ancestral configurations that are possible for a given tree size. Indeed, the next section investigates the effect of tree balance on the number of root configurations in a tree. Fig.~\ref{smalltrees} suggests that for random labeled topologies of a specified size, we can expect the variance of the number of root configurations to be large. We will confirm this claim in Section~\ref{trandom}. We will also show that, although there exist tree families (e.g.~caterpillars) for which the growth of the number of root configurations is polynomial in the tree size, the expected number of root configurations in a random labeled topology of given size $n$ grows exponentially in $n$.

\section{Root configurations for unbalanced and balanced families of trees}
\label{families}

\fil{In this section, we study the number of root configurations for particular families of trees, extending beyond two cases considered by \cite{Wu12}: the caterpillar case, which was studied exactly, and the completely balanced case, for which a loose lower bound of $(\sqrt{2})^n$ was reported.} As balance is an important tree property that influences ancestral configurations, we study unbalanced and balanced families generated by different seed trees. Upper and lower bound results on the number of root configurations for trees of specified size appear in Section~\ref{smalarge}.

\begin{figure}
\begin{center}
\includegraphics*[scale=0.80,trim=0 0 0 0]{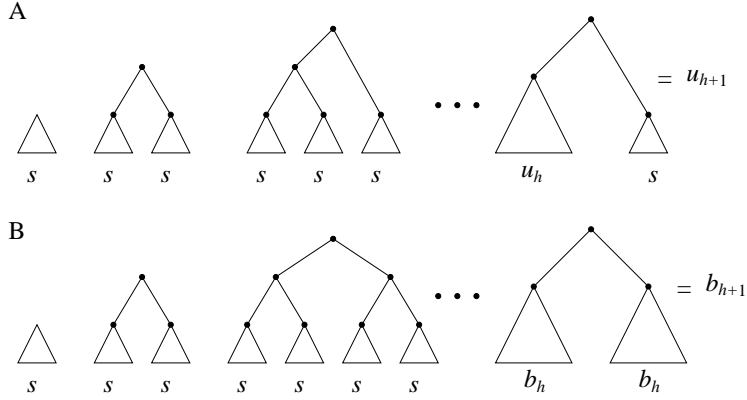}
\end{center}
\vspace{-.7cm}
\caption{{\small Unbalanced and balanced families of trees defined from a given seed tree $s$. {\bf (A)} The unbalanced family $u_h=u_h(s)$ is defined by $u_0=s$, setting $u_{h+1}$ as the tree of size $|u_{h+1}|=|u_h|+|s| =(h+2)|s| $ obtained by appending $u_h$ and $s$ to a shared root node.  {\bf (B)} The balanced family $b_h=b_h(s)$ is defined by $b_0=s$, setting $b_{h+1}$ as the tree of size $|b_{h+1}| = 2|b_h| = 2^{h+1}|s|$ obtained by appending two copies of $b_h$ to a shared root node.
}} \label{famiglie}
\end{figure}

For a given seed tree $s$, we consider the unbalanced family $(u_h(s))$ (Fig.~\ref{famiglie}A) and the balanced family $(b_h(s))$ (Fig.~\ref{famiglie}B) defined as follows
\begin{eqnarray} \label{uh}
u_0(s) &=& s\, ; \,\, u_{h+1}(s) = (u_h(s) , s )  \\ \label{bh}
b_0(s) &=& s \,; \,\, b_{h+1}(s) = (b_h(s) , b_h(s)  ), 
\end{eqnarray}
where $(t_1,t_2)$ is the tree shape obtained by appending trees $t_1$ and $t_2$ to a shared root node. Note that
the family of caterpillar trees is obtained as $(u_h(s))$, when $|s|=1$. For the same seed tree of size $1$, $(b_h(s))$ is the family of completely balanced trees. When $|s|=2$, $(u_h(s))$ resembles the lodgepole \citep{DisantoAndRosenberg15} family (the difference from \cite{DisantoAndRosenberg15} being only that here, each leaf is in a cherry, whereas in \cite{DisantoAndRosenberg15}, there is a unique leaf that is not in a cherry). For each family, it is understood that we consider an arbitrary labeling of each unlabeled shape in the family.


\subsection{Unbalanced families}

Fix a seed tree $s$ and consider the family $u_h = u_h(s)$ as defined in (\ref{uh}). Let $\gamma = \gamma_0$ be the number of root configurations in $s=u_0$, and define $\gamma_h$ as the number of root configurations in $u_h$.  If $s$ is the 1-taxon tree, then as noted earlier, the number of root configurations $\gamma$ is set to $0$.
From Proposition~\ref{prop1}, we obtain the recursion
\begin{equation}\label{pippous}
\gamma_{h+1}= 1 + \gamma + \gamma_h(\gamma + 1),
\end{equation}
starting with $\gamma_0 = \gamma$. As shown in Appendix 2, the generating function
$$U_{\gamma}(z)= \sum_{h=0}^{\infty} \gamma_h z^h$$
is described by
\begin{equation}\label{pippo}
U_{\gamma}(z) = \frac{ z + \gamma }{ (1-z)(1-z-\gamma z)}.
\end{equation}
For $\gamma \geq 0$, the dominant singularity of $U_{\gamma}$---the singularity nearest to the origin---is the solution $z_0=1/(\gamma + 1) \leq 1$ of the equation $1-z-\gamma z = 0$. Applying Theorem~IV.7 of \cite{FlajoletAndSedgewick09} yields the exponential growth of the sequence $(\gamma_h)$ with respect to the index $h$ as
\begin{equation}\label{qw}
\gamma_h \bowtie \left(\frac{1}{z_0}\right)^h = (\gamma+1)^h.
\end{equation}
Because $u_h$ has $|u_h|=(h+1)|s|$ leaves,
substituting $h = |u_h|/|s| - 1$ in (\ref{qw}), we obtain the next proposition.
\begin{prop}\label{scur}
In the unbalanced family $(u_h)$, the exponential growth of the number of root configurations in the size $|u_h|$ is
\begin{equation}\label{su}
\big[(\gamma+1)^{1/|s|}\big]^{|u_h|},
\end{equation}
where $|s|$ is the size of the seed tree and $\gamma$ is its number of root configurations. The total number of configurations in the family $(u_h)$ has the same exponential growth.
\end{prop}
In other words, for values of the number of leaves $n$ at which a member of the unbalanced family exists, the number of root configurations in the unbalanced family grows with $\big[(\gamma+1)^{1/|s|}\big]^n$.

When the seed tree is the 1-taxon tree, so that $\gamma = 0$ and $(u_h)$ is the sequence of caterpillar trees, formula (\ref{su}) gives the exponential growth $1^{|u_h|} = 1$. Indeed, the number of root configurations in the caterpillar family grows like a polynomial function of the size, as immediately follows from (\ref{pippous}) (see also \cite{Wu12}). Taking $|s|>1$, the number of root configurations in $u_h(s)$ becomes exponential in the tree size. Table \ref{tavola1} illustrates that for unbalanced families defined by small seed trees \fil{of size greater than one}, root configurations in $n$-taxon trees---provided that a tree with $n$ taxa is in the family---have exponential growth in the range $1.3^n$ to $1.5^n$.

\subsection{Balanced families}
\label{balfam}
The results change when we consider balanced families. For a fixed seed tree $s$, consider the family $b_h = b_h(s)$ as defined in (\ref{bh}). Let $\gamma = \gamma_0$ be the number of root configurations in seed tree $s=b_0$, and define $\gamma_h$ as the number of root configurations in $b_h$.  If $|s|=1$, then $\gamma$ is $0$. From Proposition~\ref{prop1}, we obtain
\begin{equation*}\label{em}
\gamma_{h+1}= (\gamma_h+1)^2,
\end{equation*}
with $\gamma_0 = \gamma$.
Defining the sequence $x_{h+1} = x_h^2 + 1$, with $x_0 = \gamma + 1$, it is straightforward to show that $\gamma_h=x_h-1$.

Sequence $x_h$ can be studied as in \cite{AhoAndSloane73} (Section~3 and Example 2.2). For $h \geq 1$, a constant $k_{\gamma}$ exists for which
\begin{equation*}
x_h = \lfloor k_{\gamma}^{(2^h)} \rfloor,
\end{equation*}
where $\lfloor \cdot \rfloor $ is the floor function. The constant $k_{\gamma}$ can be approximated using the recursive definition of $x_h$, summing terms in a series:
\begin{equation}\label{eqkappa}
k_{\gamma} = (\gamma+1) \exp\left[ \sum_{h=0}^\infty 2^{-h-1} \log\left(1 + \frac{1}{x_h^2} \right) \right].
\end{equation}
Switching back to $\gamma_h$, for $h\geq 1$, we obtain
\begin{equation}\label{rus}
\gamma_h = x_h - 1 = \lfloor k_{\gamma}^{(2^h)} \rfloor - 1.
\end{equation}
Thus, because $\gamma_h$ grows with $\lfloor k_{\gamma}^{(2^h)} \rfloor$, to determine the exponential growth of the number of root configurations, it remains to evaluate the constant $k_{\gamma}$. Rescaling (\ref{rus}) to consider the number of leaves $|b_h| = 2^h |s|$ as a parameter, we obtain the next proposition.
\begin{prop}\label{reggia}
In the balanced family $(b_h)$, the exponential growth of the number of root configurations in the size $|b_h|$ is
\begin{equation}\label{san}
\big[ (k_{\gamma})^{1/|s|} \big]^{|b_h|},
\end{equation}
where $|s|$ is the size of the seed tree. The constant $k_{\gamma}$ can be computed as in (\ref{eqkappa}) and bounded by
\begin{equation}\label{male}
\gamma+1 < k_{\gamma} < (\gamma + 1) + \frac{1}{\gamma + 1}.
\end{equation}
The total number of configurations in the family $(b_h)$ has the same exponential growth.
\end{prop}

\smallskip
\noindent In other words, for values of the number of leaves $n$ at which a member of the balanced family exists, the number of root configurations in the balanced family grows with $\big[ (k_{\gamma})^{1/|s|} \big]^n$.

\smallskip
\emph{Proof.} It remains only to prove the bound (\ref{male}). The lower bound follows quickly from (\ref{eqkappa}), as the exponent is positive. The upper bound is obtained by observing that the sequence $x_h = x_{h-1}^2+1$ is increasing, and thus $\log(1 + 1/x_0^2) \geq \log(1 + 1/x_h^2)$ for each $h \geq 0$. Therefore, from (\ref{eqkappa}) and the fact that $x_0 = \gamma + 1$, we have
$$k_{\gamma} < (\gamma+1) \exp \bigg[ \log \bigg(1 + \frac{1}{x_0^2} \bigg) \sum_{h=0}^\infty 2^{-h-1} \bigg] = (\gamma+1) \bigg[1+ \frac{1}{(\gamma+1)^2} \bigg]. \quad \Box $$

Comparing the number of root configurations in balanced families to those in unbalanced families (Table \ref{tavola1}), we see that the exponential order for balanced families is greater than in unbalanced families, though typically still in the range $1.3^n$ to $1.5^n$.

\begin{table}
\vspace{-.2cm}
\caption{{\small \fil{Approximate values of} the constants that when raised to the power $n$ describe the exponential growth with the number of taxa $n$ of the number of ancestral configurations, in unbalanced and balanced families for small seed trees}}
\label{tavola1}
\fontsize{9}{11}\selectfont
\begin{center}
\begin{tabular}{| c | c | c | c | c ||| c | c | c | c | c |}\hline \label{tavola}
 &  &  &  &  &  &  &  &  &  \\
Seed tree $s$ & $|s|$ & $\gamma$ & $(\gamma+1)^{1/|s|}$ & $(k_{\gamma})^{1/|s|}$ & Seed tree $s$ & $|s|$ & $\gamma$ & $(\gamma+1)^{1/|s|}$ & $(k_{\gamma})^{1/|s|}$  \\
  &  &  &  (unbalanced) & (balanced) & &  &  & (unbalanced) & (balanced)  \\[1.8ex] \hline
&  &  &  &  &  &  &  &  &       \\
\includegraphics[scale=.18,bb=200 200 210 250]{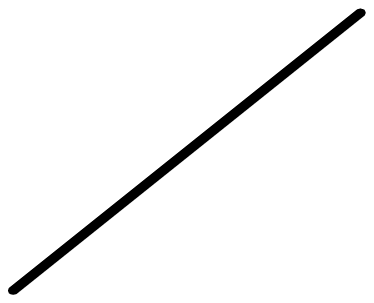}       & 1 & 0  & 1     & 1.503
& \includegraphics[scale=.18,bb=200 200 210 250]{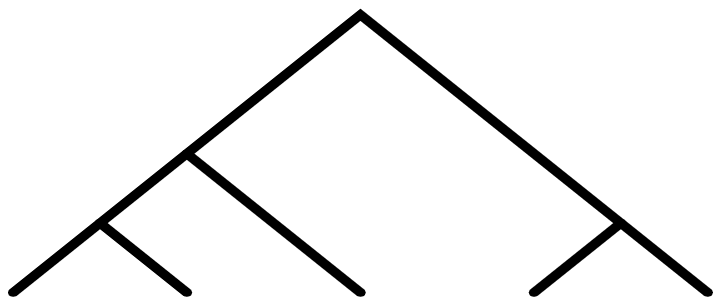}   & 5 & 6  & 1.476 & 1.479   \\[1.8ex]
\includegraphics[scale=.18,bb=200 200 210 250]{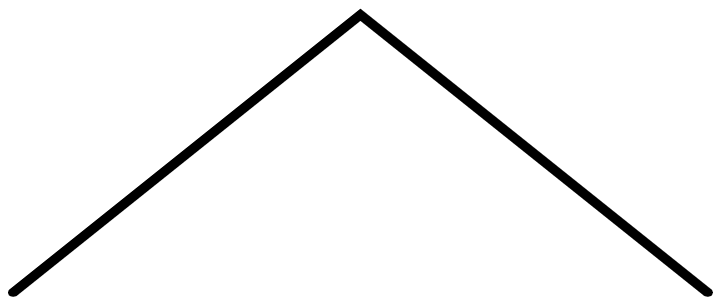}       & 2 & 1  & 1.414 & 1.503
& \includegraphics[scale=.18,bb=200 200 210 250]{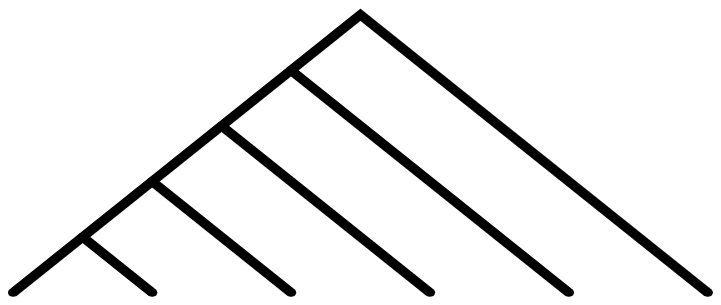}   & 6 & 5  & 1.348 & 1.351  \\[1.8ex]
\includegraphics[scale=.18,bb=200 200 210 250]{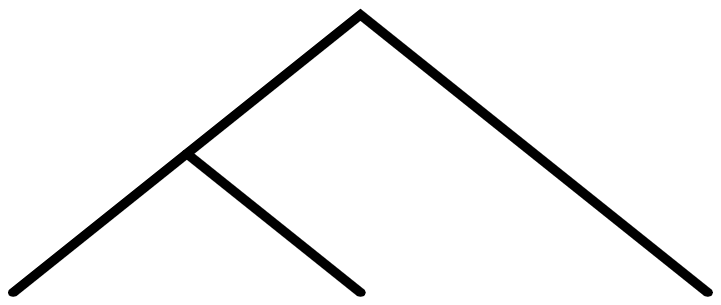}       & 3 & 2  & 1.442 & 1.469
& \includegraphics[scale=.18,bb=200 200 210 250]{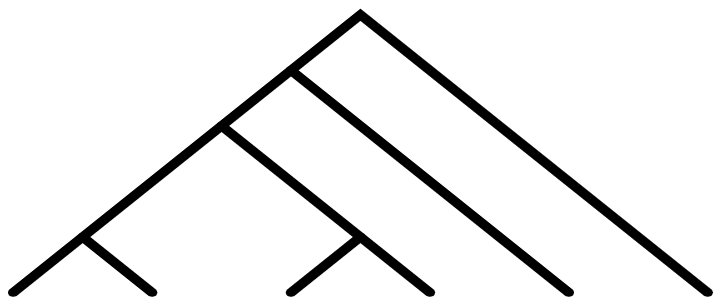}   & 6 & 6  & 1.383 & 1.385  \\[1.8ex]
\includegraphics[scale=.18,bb=200 200 210 250]{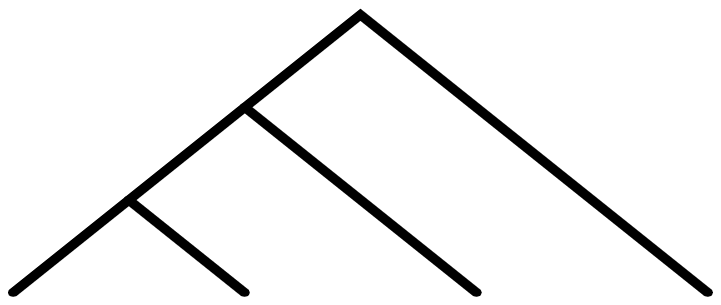}     & 4 & 3  & 1.414 & 1.425
&  \includegraphics[scale=.18,bb=200 200 210 250]{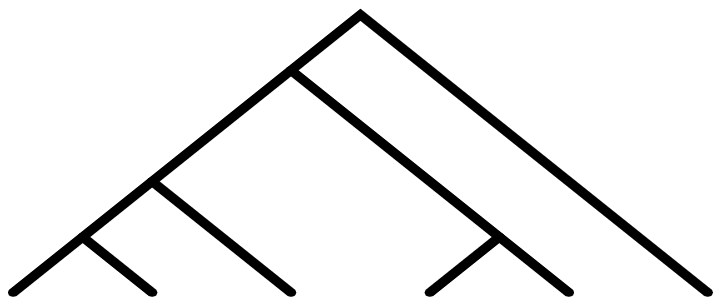}  & 6 & 7  & 1.414 & 1.416 \\[1.8ex]
\includegraphics[scale=.18,bb=200 200 210 250]{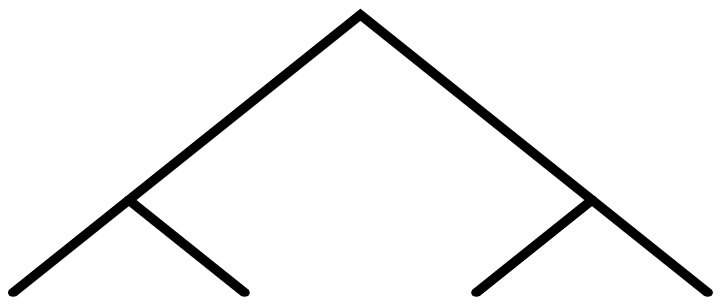}     & 4 & 4  & 1.495 & 1.503
&  \includegraphics[scale=.18,bb=200 200 210 250]{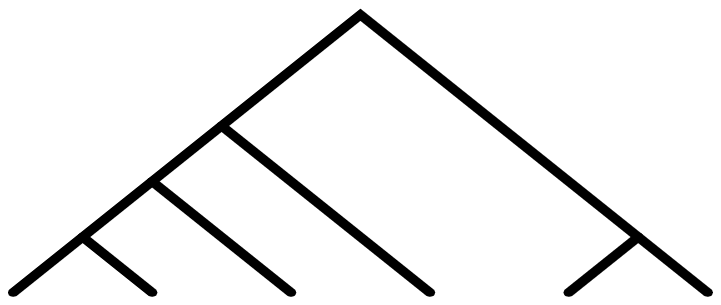}  & 6 & 8  & 1.442 & 1.444  \\[1.8ex]
\includegraphics[scale=.18,bb=200 200 210 250]{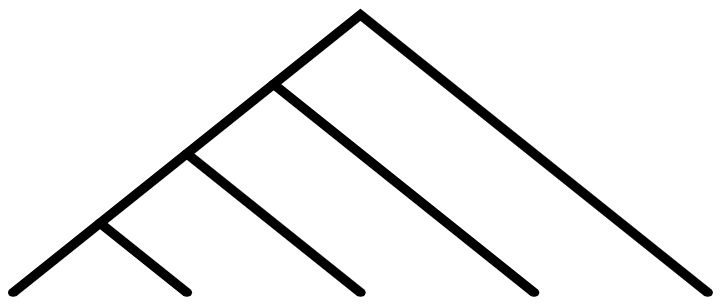}     & 5 & 4  & 1.380 & 1.385
&   \includegraphics[scale=.18,bb=200 200 210 250]{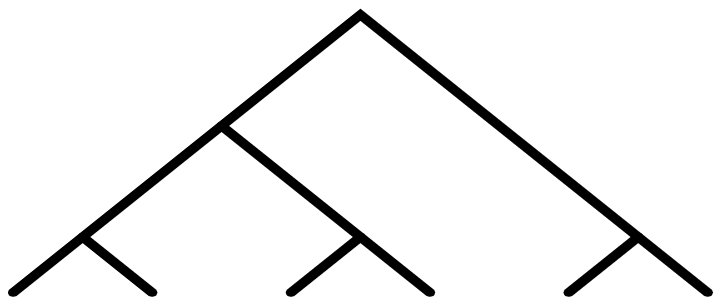} & 6 & 10 & 1.491 & 1.492   \\[1.8ex]
\includegraphics[scale=.18,bb=200 200 210 250]{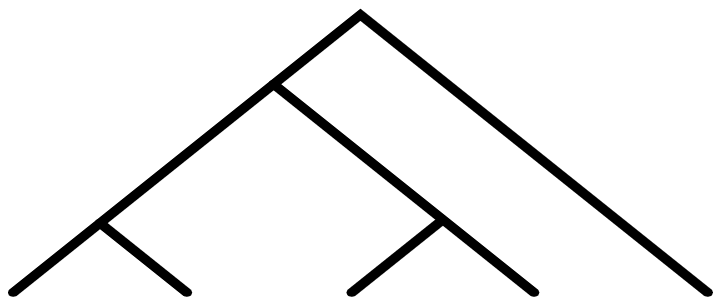}     & 5 & 5  & 1.431 & 1.435
&  \includegraphics[scale=.18,bb=200 200 210 250]{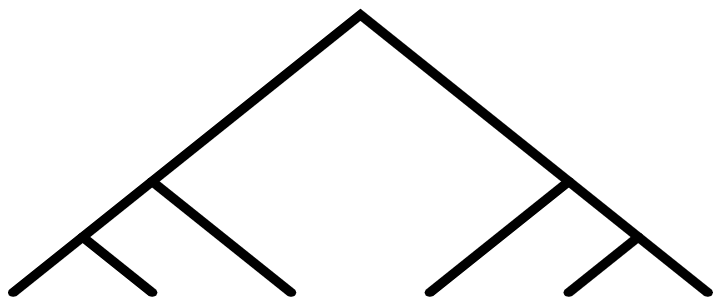}  & 6 & 9  & 1.468 & 1.469  \\[1.8ex]
\hline
\end{tabular}
\end{center}
\vspace{-.1cm}
Each constant is obtained to three decimal places by numerically evaluating (\ref{eqkappa}).
\end{table}


\subsection{Comparing unbalanced $u_h$ and balanced $b_h$ families}
For a given seed tree $s$, the quantities $u(s) = (\gamma+1)^{1/|s|}$ and $b(s) = (k_{\gamma})^{1/|s|}$ determine the exponential orders of the sequences considered in Propositions~\ref{scur} and \ref{reggia}, respectively. We observe three facts.

\smallskip

(i) Applying the lower bound in (\ref{male}), $\gamma+1 < k_{\gamma}$, for a fixed seed tree $s$, we always have
\begin{equation}\label{x}
u(s) < b(s).
\end{equation}
Therefore, the growth of the number of ancestral configurations in the family $b_h(s)$ is exponentially faster than the growth in the family $u_h(s)$. When $s$ is not small, however, $u(s)$ can become close to $b(s)$. For large $s$, $\gamma$ is also large. Owing to the upper bound in (\ref{male}), although $\gamma+1 < k_{\gamma}$, $k_{\gamma}$ only slightly exceeds $\gamma+1$. Further, the exponent $1/|s|$ in the expressions for $u(s)$ and $b(s)$ further reduces the difference between them.

For instance, if $s$ is the caterpillar tree with $10$ leaves, we have $ \gamma = 9$, $u(s) = (10)^{1/10} \approx 1.2589$, and $1.2589 \approx (10)^{1/10} < b(s) < (10.1)^{1/10} \approx 1.2601$. In this case, $b(s)-u(s)$ is \fil{bounded above by a constant near $10^{-3}$}. The increasing similarity of $u(s)$ and $b(s)$ is already evident in Table \ref{tavola1}, as their values for 6-taxon seed trees are substantially closer to each other than for the smaller 1-, 2- and 3-taxon seed trees.

\smallskip

(ii) The choice of the seed tree can play an important role in the relative values of $b(s)$ and $u(s)$, as taking two different seed trees can flip inequality (\ref{x}). In fact, if $s_1$ and $s_2$ are two seed trees of the same size $|s_1|=|s_2|=|s|$ for which $\gamma_1 > \gamma_2$, then
\begin{equation}
u(s_1)>b(s_2).
\end{equation}
To obtain this result, we note that \fil{$|s| \log u(s_1)=\log( \gamma_1 + 1 ) \geq \log [(\gamma_2 + 1) + 1 ] \geq \log[ (\gamma_2 + 1) + 1/(\gamma_2+1) ] > \log k_{\gamma_2}= |s| \log b(s_2)$}, where the latter inequality follows from the upper bound (\ref{male}). The result is observable in Table \ref{tavola1}, where at fixed $|s|$ of 4, 5, or 6, $u(s)$ for some of the shapes exceeds $b(s)$ for other shapes.


\smallskip

(iii) When the seed tree $s$ is chosen as the 1-taxon tree with $|s| = 1$, the constant $b(s) = k_0$ determines an upper bound for the number of root configurations that a tree of given size can have. This result is shown in more detail in the following section. The value of $k_0$ can be computed numerically from (\ref{eqkappa}):
\begin{equation}\label{gatto}
k_0 \approx 1.502836801.
\end{equation}
\fil{This constant provides the exact value for which $\sqrt{2} \approx 1.4142$, reported by \cite{Wu12}, provided a lower bound.}



\section{Smallest and largest numbers of root configurations for trees of fixed size}
\label{smalarge}

We have seen that the number of root configurations for caterpillar trees grows polynomially, and that the number of root configurations in unbalanced non-caterpillar families and balanced families grows exponentially. In the examples we have considered, the exponential growth proceeds with $1.3^n$ to $1.503^n$. We now show that the caterpillar trees have the smallest number of root configurations, and that the constant $k_0$ (\ref{gatto}) in fact provides an upper bound on the exponential growth of the number of root configurations as $n$ increases. We characterize the labeled topologies that possess the largest number of root configurations at fixed $n$.


\subsection{Smallest number of root configurations}

For the caterpillar tree of size $n$, the number of root configurations is $n-1$. We show that this value, $n-1$, is the smallest number of root configurations for a tree of size $n$.

Let $c_t(r)$ denote the number of root configurations of tree $t$.
Let $m_n(r) = \mathrm{min}_{ \{t: |t|=n\} } c_t(r) $. Suppose we have shown for each $i$ with $1 \leq i \leq n-1$ that
\begin{equation}\label{emmej}
m_i(r) = i-1.
\end{equation}
The claim clearly holds for $i=1,2,3$, for each of which the sole tree $t$ has $i-1$ root configurations.

For $n \geq 2$, we use induction to prove (\ref{emmej}) for $i=n$. Suppose $t'$ is a tree of size $n$ such that \fil{$c_{t'}(r) = m_n(r)$}. The number of root configurations of $t'$ is given by Proposition~\ref{prop1} as the product $c_{t'}(r) = [c_{t_{\ell}'}(r)+1][c_{t_{r}'}(r)+1]$, where $t_{\ell}'$ and $t_r'$ are the root subtrees of $t'$. \fil{Because $t'$ has the minimal number of root configurations,} $t_{\ell}'$ and $t_r'$ must separately possess the minimal number of root configurations among trees of their size. We can then write $c_{t_{\ell}'}(r) = m_i(r)$ and $c_{t_r'}(r) = m_{n-i}(r)$, where, without loss of generality, $i$ is a certain value with $1\leq i \leq \lfloor n/2 \rfloor$. Therefore $c_{t'}(r)$ has the form $c_{t'}(r) = [m_i(r) + 1] [m_{n-i}(r) + 1]$. It is determined from the minimum
\begin{equation}\label{idra}
m_n(r) = c_{t'}(r) = \mathrm{min}_{ \{i : 1 \leq i \leq \lfloor n/2 \rfloor \} } [m_i(r) + 1][m_{n-i}(r) + 1].
\end{equation}
Applying the inductive hypothesis (\ref{emmej}), we obtain $m_n(r) = \mathrm{min}_{ \{i : 1 \leq i \leq \lfloor n/2 \rfloor \} } i(n-i)$. In the permissible range for $i$, the product $i(n-i)$ reaches its minimum value at $i=1$, equaling $n-1$ as desired.

By induction, we have shown that $(\ref{emmej})$ holds for each $i \geq 1$. Furthermore, the fact that the product $[m_i(r) + 1][m_{n-i}(r) + 1]$ in (\ref{idra}) is minimal only at $i=1$ also demonstrates that those tree shapes of size $n$ with the smallest number of root configurations can be recursively obtained by appending the 1-taxon tree and the tree shape of size $n-1$ with the smallest number of root configurations to a shared root node. Trees resulting from this recursive construction are exactly those having a caterpillar shape.


\subsection{Largest number of root configurations}

For the largest number of root configurations, we denote $M_n(r) = \mathrm{max}_{\{t : |t| = n\} } c_t(r)$. Similarly to (\ref{idra}), we seek to identify the trees $t$ that produce the maximum in the following equation, and to evaluate that maximum:
\begin{equation}
M_n(r) = \mathrm{max}_{ \{i : 1 \leq i \leq \lfloor n/2 \rfloor \} } [M_i(r) + 1][M_{n-i}(r) + 1].
\end{equation}
Note that $M_1(r)=0$. Taking $\tilde{M}_n=M_n(r)+1$, we have the recursion $$\tilde{M}_n= 1+\mathrm{max}_{ \{i : 1 \leq i \leq \lfloor n/2 \rfloor \} } \tilde{M}_i  \tilde{M}_{n-i},$$ starting with $\tilde{M}_1=1$. The sequence $\tilde{M}_n$ was studied by \cite{deMierAndNoy12} (Theorems~1 and 2), where it was shown: (i) taking $d=d(n)$ as the power of $2$ nearest to $n/2$, we have
$\tilde{M}_n = 1 + \tilde{M}_{d} \tilde{M}_{n-d}$, so that $$M_n(r) = [M_d(r) + 1][M_{n-d}(r) + 1];$$
(ii) for all $n \geq 10$,
$k_0^{n-1/4} < \tilde{M}_n < k_0^n$, that is,
\begin{equation}\label{gino}
k_0^{n-1/4} -1 < M_n(r) < k_0^n - 1,
\end{equation}
where the constant $k_0$ has been already computed in (\ref{gatto}).
%

For small $n$, the labeled topologies with the largest numbers of root configurations appear in Fig.~\ref{maxi}. Collecting the results for the smallest and largest number of root configurations, we can state the following facts.
\begin{prop}\label{matilde}
(i) For each $n\geq 1$, the smallest number of root configurations in a tree of size $n$ is $m_n(r)=n-1$. The caterpillar tree shape of size $n$ has exactly $m_n(r)$ root configurations. (ii) For each $n\geq 10$, the largest number of root configurations in a tree of size $n$, $M_n(r)$, can be bounded as in (\ref{gino}). For $n \geq 2$, if $d=d(n)$ denotes the power of $2$ nearest to $n/2$, then $M_n(r)$ is the number of root configurations in the tree shape $t_n$ recursively defined as $|t_1| = 1$, $t_n = (t_d,t_{n-d})$. When $n=2^h$ for integers $h$, $t_n$ is the completely balanced tree of depth $h$, and $M_n(r) = \lfloor k_0^n \rfloor - 1$ (\ref{rus}).
\end{prop}
As a corollary, we obtain the following result, the proof of which appears in Appendix 3.
\begin{coro}\label{corollario}
$(i)$ The exponential growth of the sequences $m_n(r)= \mathrm{min}_{ \{t:|t|=n \}} c_t(r)$ and $M_n(r)= \mathrm{max}_{ \{t:|t|=n \}} c_t(r)$ follows $m_n(r) \bowtie 1$ and $M_n(r) \bowtie k_0^n$. $(ii)$ The sequences $m_n = \mathrm{min}_{ \{t:|t|=n \}} c_t$ and $M_n = \mathrm{max}_{ \{t:|t|=n \}} c_t$, giving respectively the smallest and the largest total number of configurations $c_t$ in a tree $t$ of size $n$, have exponential growth $m_n \bowtie m_n(r)$ and $ M_n \bowtie M_n(r)$.
\end{coro}

\begin{figure}
\begin{center}
\includegraphics*[scale=0.9,trim=0 0 0 0]{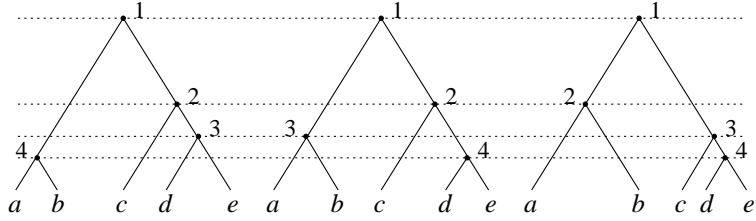}
\end{center}
\vspace{-.7cm}
\caption{{\small The three labeled histories of the labeled topology $t=((a,b),(c,(d,e)))$ of size $n=5$. Each labeled history can be represented by bijectively labeling the $n-1$ internal nodes of $t$ with the integers in $[1,n-1]$ in such a way that each path from the root of $t$ to a leaf of $t$ is labeled by an increasing sequence.}} \label{histo}
\end{figure}

The family of tree shapes $(t_n)$ defined in Proposition~\ref{matilde} by the recursive decomposition $|t_1|=1$ and $t_n=(t_d,t_{n-d})$, where $d$ is the power of $2$ nearest to $n/2$, already has a place in the study of gene trees and species trees, as it provides the ``maximally probable'' tree shapes of \cite{DegnanAndRosenberg06}. Given a labeled topology $t$ of size $n$, a labeled history of $t$ is a linear ordering of the $n-1$ internal nodes of $t$ such that the order of the nodes in each path going from the root of $t$ to a leaf of $t$ is increasing (Fig.~\ref{histo}). As reported by \cite{Harding74} and proved by \cite{HammersleyAndGrimmett74}, each labeled topology with $t_n$ as its underlying unlabeled topology possesses the maximal number of labeled histories among labeled topologies of size $n$. Consider the Yule model for the probability distribution of tree shapes, in which pairs of lineages in a labeled set of $n$ lineages are joined together, at each step choosing uniformly among pairs \citep{Yule25, Harding71, Brown94, McKenzieAndSteel00, SteelAndMcKenzie01, Rosenberg06:anncomb, DisantoEtAl13, DisantoAndWiehe13}. Among all labeled topologies with size $n$, those with the largest number of labeled histories---and hence, with shape $t_n$---have the highest probability under the model.

For $n \geq 3$, the maximally probable labeled topologies of size $n$---those with the most labeled histories---can be recursively characterized as those labeled topologies whose two root subtrees are maximally probable labeled topologies of sizes $s=2^k$ and $n-s$, where $k = 1 + \lfloor \log_2[(n-1)/3] \rfloor$ \citep{HammersleyAndGrimmett74, Harding74}. This characterization matches our characterization that the unlabeled shapes with the largest number of root configurations are those for which the subtrees have the most root configurations and sizes $d$ and $n-d$, where $d=2^\ell$ is the nearest power of 2 to $n/2$.

To see that the characterizations are identical so that $\{d,n-d\}=\{s,n-s\}$, note that a specific $2^\ell$ is the nearest power of 2 to $n/2$ precisely for integers $n \in [2^\ell + 2^{\ell-1},2^\ell + 2^{\ell+1}] = [3 \times 2^{\ell-1}, 3 \times 2^\ell]$. On the endpoints of the interval, there are two choices for $d$, but in both cases, one choice is $2^\ell$. At the same time, the integers $n$ for which $\ell = 1 + \lfloor \log_2[(n-1)/3 ] \rfloor$ are precisely those in $[3 \times 2^{\ell-1} + 1, 3 \times 2^\ell]$. Thus, $\{s,n-s\} = \{d,n-d\}$ for all integers $n$ in $[3 \times 2^{\ell-1} + 1, 3 \times 2^\ell]$. On the lower boundary, for $n = 3 \times 2^{\ell-1}$, $s=2^{\ell -1}$ and $\{s,n-s\} = \{2^{\ell -1}, 2^\ell \} = \{d,n-d\}$. Dividing the integers in $[3, \infty )$ into a union of intervals $\cup_{\ell=1}^\infty [3 \times 2^{\ell - 1}, 3 \times 2^\ell)$, we see that $\{s,n-s\} = \{d,n-d\}$ on each interval, and hence, $\{s,n-s\} = \{d,n-d\}$ for all $n \geq 3$.

\begin{figure}
\begin{center}
\includegraphics*[scale=0.5,trim=0 0 0 0]{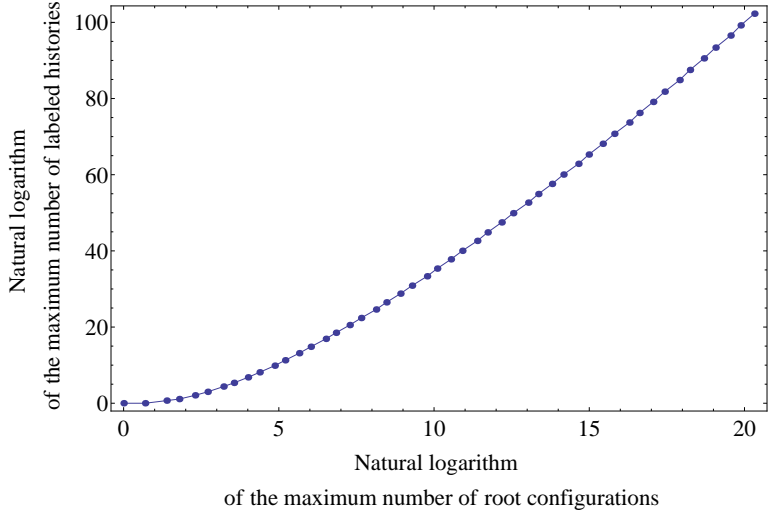}
\end{center}
\vspace{-.7cm}
\caption{{\small Natural logarithm of the maximum number of histories possible for a labeled topology of size $n$ as a function of the natural logarithm of the maximum number of root configurations possessed by a labeled topology of the same size ($2\leq n \leq 50$). The maxima occur at the same set of labeled topologies.}} \label{confiHist}
\end{figure}

This result shows that for a given tree size, those labeled topologies whose shapes belong to the family $(t_n)$ maximize both the number of root configurations and the number of labeled histories. For these labeled topologies, in Fig.~\ref{confiHist}, we plot the logarithm of the maximum number of labeled histories possible for a labeled topology of size $n$ as a function of the logarithm of the maximum number of root configurations. Although the shapes are the same, the number of labeled histories is considerably larger than the number of root configurations. The growth is approximately linear, suggesting that the maximal number of labeled histories increases approximately exponentially in the maximal number of root configurations.


\section{The number of root configurations in a random labeled topology}
\label{trandom}
We now study through generating functions the number of root configurations when trees of a given size are randomly selected under a uniform distribution on the set of labeled topologies. In Section~\ref{media}, we show that
the expectation $\mathbb{E}_n[c(r)]$ of the number of root configurations in a random labeled topology of size $n$ has exponential growth $(4/3)^n$. In Section~\ref{varia}, we show that the variance $\mathrm{Var}_n[c(r)]$ of the number of root configurations has exponential growth $[ \frac{4}{7(8\sqrt{2}-11)} ]^n$. The same results hold for the random total number of configurations.

\subsection{Mean number of root configurations}\label{media}

Define the exponential generating function
\begin{equation}
\label{eqFz}
F(z) = \sum_{t \in T} \frac{c_t(r)}{|t|!} z^{|t|},
\end{equation}
where $c_t(r)$ is the number of root configurations in tree $t$. As shown in Appendix 4, the function $F$ satisfies
\begin{equation}\label{equaf}
F(z) = \frac{1}{2} [F(z) + T(z)]^2,
\end{equation}
where $T(z)$ is the exponential generating function in (\ref{expo}).
Solving (\ref{equaf}), we obtain a closed form for $F(z)$,
\begin{equation}\label{giunio}
F(z) = \sqrt{1-2z} -\sqrt{ 2 \sqrt{1-2z} -1  } = \frac{z^2}{2} + z^3 + 2z^4 + \frac{33z^5}{8} + \ldots
\end{equation}
Here we have taken the negative root of the quadratic equation, as it is the root that produces the correct value of $F(z)=0$ at $z=0$. \fil{It can be seen that $F(0)=0$ is required by noting that the first term in the expansion (\ref{eqFz}) of $F(z)$ is the $z^1$ term, as the set $T$ contains only trees of size at least 1, so that (\ref{eqFz}) has no constant term.}

The value of $z$ that cancels the second square root in (\ref{giunio}) is $z=3/8$, which is smaller than the value $z=1/2$ that cancels the first square root, $\sqrt{1-2z}$. In the complex plane, both $z=3/8$ and $z=1/2$ are singularities of $F(z)$. The dominant singularity is $z=3/8$, as it is nearer to the origin. To highlight the type of singularity that $F(z)$ has at the point $z=3/8$, it is convenient to factor the second square root in (\ref{giunio}), writing $F(z)$ as
$$F(z) = \sqrt{1-2z} -\bigg( \sqrt{1-\frac{8}{3}z} \bigg) f(z),$$
where
\begin{equation}
f(z) = \sqrt{ \frac{3(1-2\sqrt{1-2z})}{8z-3}  }
\end{equation}
is an analytic function in the circle $\{z \in \mathbb{C}: |z|< 1/2   \}$, except at a removable singularity $f(3/8) = \sqrt{3/2}$. Thus, we see that at $z=3/8$, the generating function $F(z)$ has a singularity of the square root type.

We can then apply Theorems~VI.1 and VI.4  of \cite{FlajoletAndSedgewick09} to recover the asymptotic behavior of the $n$th coefficient of $F(z)$,
$$[z^n]F(z) = \frac{1}{n!}\sum_{t \in T_n} c_t(r), $$
as the $n$th coefficient of the expansion of $F(z)$ at the singularity $z=3/8$. This expansion is given by
$$F(z) = \sqrt{1-2 \frac{3}{8}} - \bigg( \sqrt{1-\frac{8}{3}z} \bigg)  f(3/8) = \frac{1}{2} - \sqrt{1-\frac{8}{3}z}  \sqrt{\frac{3}{2}}.$$
We thus have
\begin{equation}
\frac{1}{n!} \sum_{t \in T_n} c_t(r) \sim [z^n]\left(-\sqrt{1-\frac{8}{3}z} \right) \sqrt{\frac{3}{2}} \sim \frac{1}{2\sqrt{\pi n^3}} \left( \frac{8}{3} \right)^n \sqrt{\frac{3}{2}}   \sim \sqrt{\frac{3}{8}} \frac{1}{\sqrt{\pi n^3}} \left( \frac{8}{3} \right)^n,
\end{equation}
where we have used the asymptotic relation $[z^n](-\sqrt{1-z}) \sim 1/(2\sqrt{\pi n^3})$ \citep{FlajoletAndSedgewick09}. Dividing by the number of trees of size $n$, $|T_n|$, as given in (\ref{carciofo}), using Stirling's formula $n! \sim \sqrt{2\pi n}(n/e)^n$, and noting the definition of $\mathbb{E}_n[c(r)]$ as a mean over all labeled topologies, we obtain the asymptotic expected number of root configurations in a random labeled topology of size $n$:
\begin{equation}\label{exp}
\mathbb{E}_n[c(r)] = \frac{\sum_{t \in T_n} c_t(r)}{|T_n|} \sim \frac{\sqrt{\frac{3}{8}} \frac{1}{\sqrt{\pi n^3}} \left( \frac{8}{3} \right)^n n! }{ \frac{(2n)!}{(2n-1) 2^n n!}   } \sim \sqrt{\frac{3}{2}} \left( \frac{4}{3} \right)^n.
\end{equation}

We summarize these results in a proposition.
\begin{prop}\label{gennarino}
The mean number of root configurations in a random labeled topology of size $n$ among the $|T_n|$ labeled tree topologies is asymptotically
$$\mathbb{E}_n[c(r)] \sim \sqrt{\frac{3}{2}} \left( \frac{4}{3} \right)^n.$$
The mean total number of configurations has exponential growth $$\mathbb{E}_n[c] \bowtie \mathbb{E}_n[c(r)] \bowtie (4/3)^n.$$
\end{prop}

\begin{figure}
\begin{center}
\begin{tabular}{c c}
\includegraphics*[scale=0.5,trim=0 0 0 0]{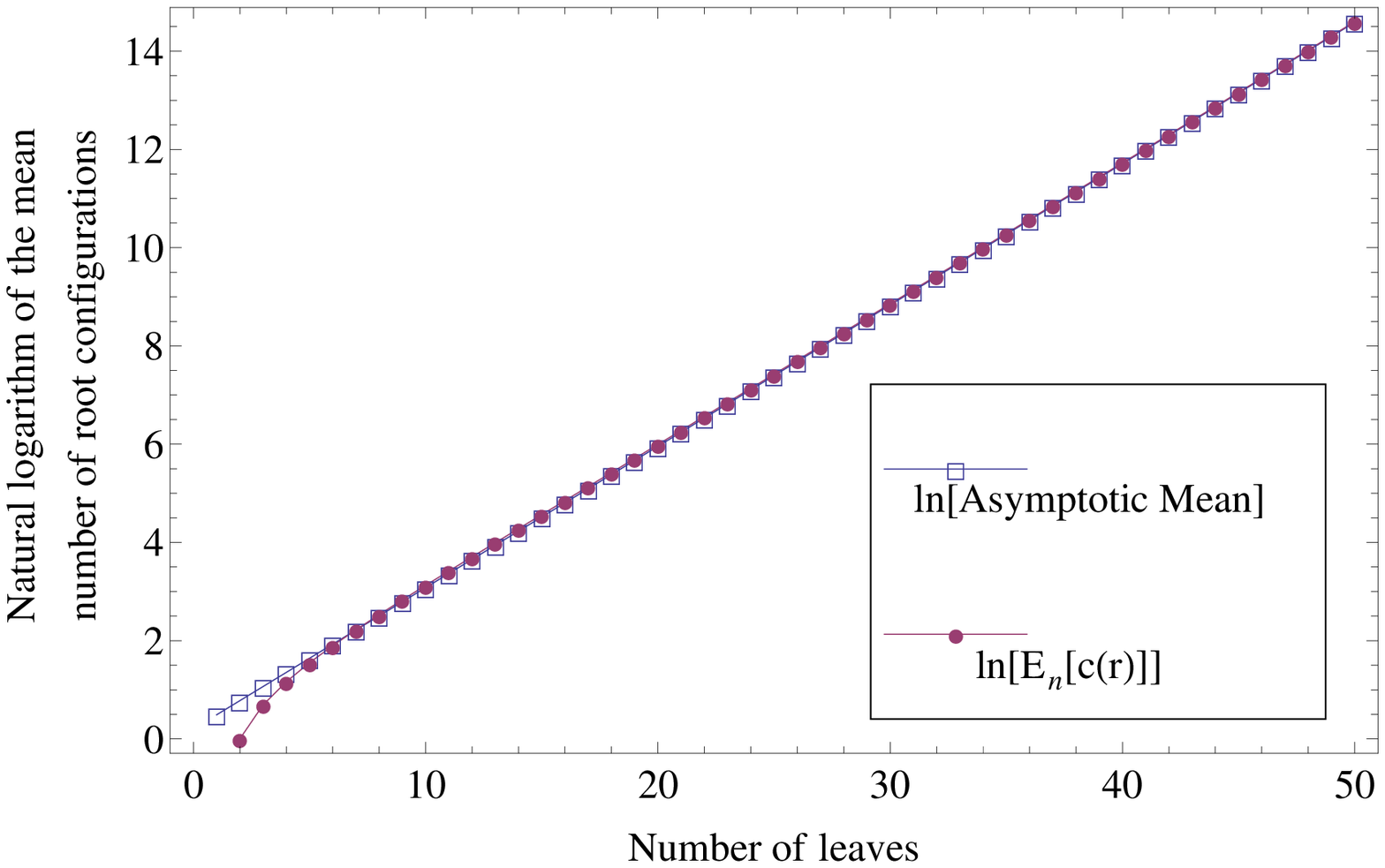} &
\includegraphics*[scale=0.5,trim=0 0 0 0]{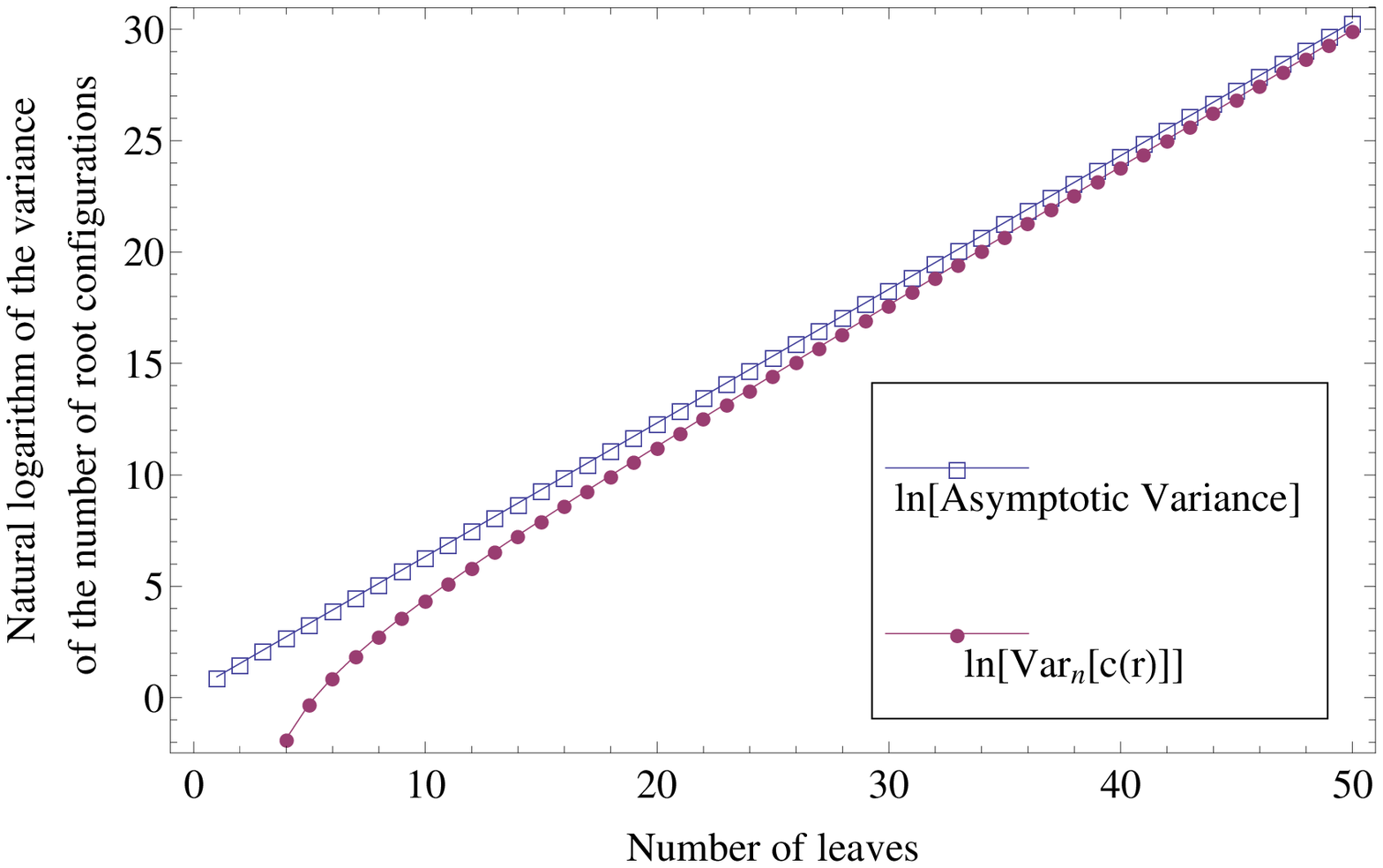} \\
\end{tabular}
\end{center}
\vspace{-.7cm}
\caption{{\small Mean and variance of the number of root configurations in random labeled topologies of fixed size. {\bf (A)} Exact natural logarithm of the mean, computed from the power series expansion of $F(z)$ (\ref{giunio}), and its asymptotic approximation from Proposition \ref{gennarino}. {\bf (B)} Exact natural logarithm of the variance, computed from the power series expansion of $G(z)$ (\ref{pallone}), and its asymptotic approximation from Proposition \ref{gennarina}.
}} \label{figPalazzo}
\end{figure}

\smallskip In Fig.~\ref{figPalazzo}A, we can see that the approach of the \fil{natural logarithm of the} exact mean number of root configurations---computed by evaluating the expansion of the generating function $F(z)$---to the asymptotic value \fil{$\log [\sqrt{3/2} (4/3)^n]$} proceeds quickly, so that even with small values of $n$, the exact mean and the asymptote are quite close \fil{on a logarithmic scale}.

\subsection{Variance of the number of root configurations}
\label{varia}
By applying the same approach used to determine the mean value of the number of root configurations across labeled topologies, in this section, we study the expectation $\mathbb{E}_n[c(r)^2]$ and then derive the asymptotic variance $\mathrm{Var}_n[c(r)]$ of the number of root configurations.

Define the generating function
\begin{equation}
G(z) = \sum_{t\in T} \frac{c_t(r)^2}{|t|!}z^{|t|}.
\end{equation} As shown in Appendix 5, the function $G(z)$ satisfies
\begin{equation}\label{cappuccino}
G(z) = \frac{1}{2} [G(z) + 2F(z) + T(z)]^2.
\end{equation}
This equation relates $G(z)$ to the generating functions $F(z)$ and $T(z)$ appearing in (\ref{giunio}) and (\ref{expo}). Solving for $G(z)$, we obtain the function
\begin{eqnarray}\label{pallone}
G(z) &=& -\sqrt{4 \sqrt{ 2\sqrt{1-2z}-1} - 2\sqrt{1-2z}-1} + 2\sqrt{2\sqrt{1-2z}-1} - \sqrt{1-2z} \\ \nonumber
&=& \frac{z^2}{2} + 2z^3 + \frac{13z^4}{2} + \frac{161z^5}{8}+ \ldots,
\end{eqnarray}
which has its dominant singularity at $z =  7(8\sqrt{2}-11)/8 \approx 0.2745 < 3/8 < 1/2$. Here, in the same way as in the derivation of $F(z)$, we have taken the negative root of the quadratic equation (\ref{cappuccino}), as it is this root that produces the correct value of $G(z)=0$ at $z=0$. At the dominant singularity for $z$, the first square root in (\ref{pallone}) cancels. Factoring this square root, the function $G(z)$ can be written as
$$G(z) = - \bigg[\sqrt{1-\frac{8z}{7(8\sqrt{2}-11)} } \bigg] g(z)+2\sqrt{2\sqrt{1-2z}-1} - \sqrt{1-2z},$$
where
\begin{equation}\label{pigna}
g(z) = \sqrt{ \frac{ 49(1-4\sqrt{2\sqrt{1-2z}-1} + 2\sqrt{1-2z} )}{8(11+8\sqrt{2})z-49} }.
\end{equation}
The function $g(z)$ is analytic in the circle $\{z \in \mathbb{C}: |z| < 3/8 \}$, except at the removable singularity $ g(7(8\sqrt{2}-11)/8) = 1.4048 \ldots$. By Theorems~VI.1 and VI.4 of \cite{FlajoletAndSedgewick09}, we can recover the asymptotic behavior of the $n$th coefficient $z^n[G(z)] = (1/n!)\sum_{t \in T_n} c_t(r)^2$ as
\begin{equation}
\frac{1}{n!}\sum_{t \in T_n} c_t(r)^2 \sim [z^n]\left(  - \sqrt{1-\frac{8z}{7(8\sqrt{2} - 11)} }\right) g(7(8\sqrt{2}-11)/8  )  \sim \frac{g(7(8\sqrt{2}-11)/8  )}{2\sqrt{\pi n^3}} \left[ \frac{8}{7(8\sqrt{2}-11)} \right]^n.
\end{equation}
Dividing by $|T_n|$ and using Stirling's approximation yields
\begin{equation}
\mathbb{E}_n[c(r)^2] = \frac{\sum_{t \in T_n} c_t(r)^2}{|T_n|}  \sim g(7(8\sqrt{2}-11)/8  ) \left[ \frac{4}{7(8\sqrt{2}-11)} \right]^n.
\end{equation}

To obtain an asymptotic estimate for the variance, we use (\ref{exp}) to note that the exponential growth of $(\mathbb{E}_n[c(r)])^2$ is $[(4/3)^2]^n$. Because $(4/3)^2 < 4/[7(8\sqrt{2} - 11)]$, we have that as $n \rightarrow \infty$,
\begin{equation}\label{nonso}
\frac{\mathrm{Var}_n[c(r)]}{ \mathbb{E}_n[c(r)^2]} = \frac{ \mathbb{E}_n[c(r)^2] - \big(\mathbb{E}_n[c(r)]\big)^2 }{ \mathbb{E}_n[c(r)^2]  } = 1 - \frac{ \big( \mathbb{E}_n[c(r)] \big)^2  }{ \mathbb{E}_n[c(r)^2]  } \rightarrow 1,
\end{equation}
and thus, the variance asymptotically satisfies
$\mathrm{Var}_n[c(r)] \sim \mathbb{E}_n[c(r)^2]$.

Furthermore, because $\mathbb{E}_n[c] \bowtie \mathbb{E}_n[c(r)]$ and $\mathbb{E}_n[c^2] \bowtie \mathbb{E}_n[c(r)^2] $ as shown in (\ref{bau2}) and (\ref{bau3}), (\ref{nonso}) also holds when we replace $c(r)$ by $c$. Thus, the variance $\mathrm{Var}_n[c]$ of the total number of configurations in a random labeled topology of size $n$ satisfies $$\mathrm{Var}_n[c] \sim \mathbb{E}_n[c^2] \bowtie  \mathbb{E}_n[c(r)^2] \sim \mathrm{Var}_n[c(r)].$$
We summarize in a proposition.
\begin{prop}\label{gennarina}
The variance of the number of root configurations in a random labeled topology of size $n$  among the $|T_n|$ labeled tree topologies is asymptotically
$$\mathrm{Var}_n[c(r)] \sim g(7(8\sqrt{2}-11)/8  ) \left[ \frac{4}{7(8\sqrt{2}-11)} \right]^n,$$
where $g(7(8\sqrt{2}-11)/8  ) \approx 1.4048$. The variance of the total number of configurations has exponential growth $$\mathrm{Var}_n[c] \bowtie \mathrm{Var}_n[c(r)]  \bowtie  \left[ \frac{4}{7(8\sqrt{2}-11)} \right]^n .$$
\end{prop}

\smallskip Fig.~\ref{figPalazzo}B demonstrates that \fil{on a logarithmic scale,} the approach of the exact variance of the number of root configurations---\fil{computed from $(n!/|T_n|)\, z^n[G(z)] - \{(n!/|T_n|) \, z^n[F(z)]\}^2$}---to the asymptotic value $g(7(8\sqrt{2}-11)/8  ) [ \frac{4}{7(8\sqrt{2}-11)} ]^n$ occurs rapidly in $n$, though slower than was seen for the mean in Fig.~\ref{figPalazzo}A.


\section{Conclusions}

Under the assumption that the labeled gene tree topology matches the species tree topology, $G=S=t$, we have studied the number of ancestral configurations in a given phylogenetic tree $t$. In particular, we have focused on the exponential growth of the number of root configurations in $t$, a quantity that also describes the exponential growth of the total number of configurations in $t$.

In Section~\ref{families}, extending results of \cite{Wu12}, \fil{in which the the enumeration of ancestral configurations for caterpillar trees and a lower bound for their number in completely balanced trees were determined}, we considered special families of trees generated by arbitrary seed trees $s$, namely the unbalanced family $u_h(s)$ and the balanced family $b_h(s)$ (Fig.~\ref{famiglie}). The main results describing the influence of tree balance and the seed tree topology on the number of ancestral configurations are collected in Propositions~\ref{scur} and Proposition~\ref{reggia} for the unbalanced and balanced cases. We have shown that for each fixed seed tree $s$, the number of ancestral configurations in the balanced family $b_h(s)$ grows exponentially faster than in the unbalanced family $u_h(s)$. When the size of the seed tree $s$ is large, however, the difference between the exponential orders of the two integer sequences can become small. We have also observed that the choice of the seed tree can have an important influence on the number of root configurations. In fact, the number of root configurations in the family $u_h(s_1)$ can grow \fil{exponentially} faster than in the family $b_h(s_2)$ when the number of root configurations in $s_1$ exceeds that of $s_2$.

When $|s|=1$, the unbalanced family $u_h(s)$ reduces to the caterpillar family, and the balanced family $b_h(s)$ gives the family of completely balanced trees. As shown in Proposition~\ref{matilde}, among trees of size $n$, the caterpillar tree with $n$ taxa possesses the smallest number of root configurations. When $n$ is a power of $2$, the completely balanced tree of size $n$ has the largest number; more generally, the largest number of root configurations occurs at precisely the labeled topologies that for a fixed $n$ generate the largest number of labeled histories. As the caterpillar labeled topologies give rise to the smallest number of labeled histories at fixed $n$---only one---both the largest and smallest numbers of root configurations occur at trees producing the extrema in the number of labeled histories. The growth of the number of root configurations in the caterpillar family is polynomial, whereas for the completely balanced trees, it is exponential with order $k_0 \approx 1.5028$.

Assuming a uniform distribution over the labeled topologies with a given size $n$, in Section~\ref{trandom} we studied the mean and the variance of the number of ancestral configurations in a random labeled topology of size $n$. By using a generating function approach, in Propositions~\ref{gennarino} and \ref{gennarina}, we have shown that the mean number of ancestral configurations has exponential growth  $(4/3)^n$, whereas for the variance we have $\big[\frac{4}{7(8\sqrt 2 -11)} \big]^n \approx 1.8215^n$.

Our results can assist in relating the complexity of algorithms for computing gene tree probabilities based on ancestral configurations---STELLS \citep{Wu12}---to those that use an evaluation based on a different class of combinatorial objects, the ``coalescent histories" \citep{DegnanAndSalter05, Rosenberg07:jcb, RosenbergAndDegnan10, Rosenberg13:tcbb, DisantoAndRosenberg15, DisantoAndRosenberg16}. In such comparisons, we expect that the ancestral configurations will often grow slower, as is seen in comparing the polynomial growth of the number of ancestral configurations in the caterpillar case with the corresponding exponential growth of the number of coalescent histories. However, the trees with the largest numbers of coalescent histories and the trees with the largest number of ancestral configurations are not the same, so that potential exists for each type of algorithm to be favorable in different cases; it remains to be seen whether the complexity of gene tree probability calculations can be reduced by choosing the computational approach based on the tree shapes under consideration.


Many enumerative problems related to ancestral configurations remain open. For example, when ancestral configurations are grouped according to an equivalence relation defined in the appendix of \cite{Wu12} that accounts for symmetries in gene trees, the number of the resulting equivalence classes---the number of ``non-equivalent'' ancestral configurations---remains to be investigated. For gene trees and species trees with a matching labeled topology, our enumerations can be used as upper bounds for the number of non-equivalent ancestral configurations, and they can help in measuring the decrease in the number of ancestral configurations when the equivalence relation is taken into account. We defer this analysis for future work.

\subsection*{Appendix 1. Proof of (\ref{miaomiao})}
Given a tree $t$, fix without loss of generality one of the possible planar representations of the tree $t$: one of the possible drawings of $t$ in which edges do not cross and intersect only at their endpoints (Fig.~\ref{config}A).

A root configuration of $t$ uniquely determines a partition of the set of the leaves of $t$ in the following way. If $\gamma=\{k_1,k_2,\ldots,k_m \}$ is a root configuration of $t$, where each $k_i$ is a node of $t$, then the associated partition is $\gamma'=\{\ell_1,\ell_2,\ldots,\ell_m   \},$ where $\ell_i$ is the set of leaves of $t$ descended from node $k_i$ (including $k_i$ itself when $k_i$ is a leaf). For instance, the partition of the leaf label set $\{a,b,c,d,e,f\}$ associated with the root configuration $\gamma = \{a,b,\ell\}$ depicted in Fig.~\ref{config}B is $\gamma' = \{ \{a\},\{b\},\{c,d,e,f\}  \}$.
Note that for each pair of indices $i,j$ with $i \neq j$, the leaves in $\ell_i$ are either all on the left or all on the right of the leaves in $\ell_j$ in the planar representation of $t$.

Without loss of generality, we can assume that the set $\gamma'$ is indexed such that if $1\leq i < j \leq m$, then the leaves in $\ell_i$ are all depicted in the planar representation to the left of the leaves in $\ell_j$. Taking the cardinality of each element $\ell_i$ of $\gamma'$ determines the vector $\gamma''=(|\ell_1|,|\ell_2|,\ldots,|\ell_m| ),$ which represents a composition, or ordered partition, of the integer $n = |t|$. For instance, for the root configurations of the tree of size $n=6$ depicted in Fig.~\ref{config}A, we obtain the following compositions of $6$:
\begin{eqnarray}\nonumber
\{ g,\ell  \} &\rightarrow & (2,4),\,\,  \{ a,b,\ell  \} \rightarrow (1,1,4),\,\,  \{ g,h,i  \} \rightarrow (2,2,2),\,\,  \{ a,b,h,i  \} \rightarrow (1,1,2,2), \\\nonumber
\{ g,h,e,f  \} &\rightarrow & (2,2,1,1),\,\, \{ a,b,h,e,f  \} \rightarrow (1,1,2,1,1),\,\, \{ g,c,d,i  \} \rightarrow  (2,1,1,2), \\\nonumber
\{ a,b,c,d,i  \} &\rightarrow & (1,1,1,1,2),\,\,  \{ g,c,d,e,f  \} \rightarrow (2,1,1,1,1),\,\,  \{ a,b,c,d,e,f  \} \rightarrow (1,1,1,1,1,1).
\end{eqnarray}

As can be seen in this example, for a given planar representation of $t$, the mapping $\gamma \rightarrow \gamma''$ is injective (i.e.~$\gamma_1 \neq \gamma_2 \Rightarrow \gamma_1'' \neq \gamma_2''$). For $1 \leq i \leq n$, there are ${n-1 \choose i-1}$ compositions of $n$ into $i$ parts, as $i-1$ demarcations must be placed among $n-1$ possible positions between entries of the length-$n$ vector $(1,1,\ldots,1)$ to separate groups of 1's that will be aggregated together. Using the binomial theorem to sum over all possible values of $i$, the number of distinct compositions of $n$ is $\sum_{i=1}^{n} {{n-1}\choose{i-1}} = 2^{n-1}$. Because each root configuration is associated with a distinct composition of $n$, we obtain $c(r) \leq 2^{|t|-1}$, and the proof of (\ref{miaomiao}) is complete.


\subsection*{Appendix 2. Proof of (\ref{pippo})}
We obtain (\ref{pippo}) from (\ref{pippous}) by noting that for $z$ close to $0$, the following expansion holds:
\begin{eqnarray}\nonumber
U_{\gamma} (z) &=& \sum_{h=0}^\infty \gamma_h z^h = \gamma_0 + \sum_{h=0}^\infty \gamma_{h+1}z^{h+1} = \gamma + z \sum_{h=0}^\infty [1+\gamma + (1+\gamma)\gamma_h ] z^h \\\nonumber
&=& \gamma + z(1+\gamma)\sum_{h=0}^\infty z^h + z(1+\gamma)\sum_{h=0}^\infty \gamma_h z^h = \gamma + \frac{z}{1-z}(1+\gamma) + z(1+\gamma) U_{\gamma}(z).
\end{eqnarray}


\subsection*{Appendix 3. Proof of Corollary~\ref{corollario}}
The proof follows from the properties of $m_n(r)$ and $M_n(r)$ stated in Proposition~\ref{matilde}. Part (i) is immediate from  Proposition~\ref{matilde} and  the definition of the exponential order.

For (ii), we start with $m_n$. Let $m_n \bowtie k_m^n$ be the exponential growth of the sequence $m_n$, so that $k_m$ is its exponential order. Denote by $(t_n)$ the caterpillar family of trees, where $t_n$ is the caterpillar with $n\geq 1$ taxa. Thus, $c_{t_n}$ is the total number of configurations in $t_n$ and $c_{t_n}(r) = m_n(r)$ is its number of root configurations. By (\ref{doct}), we have $c_{t_n} \bowtie c_{t_n}(r)$, and $c_{t_n}(r) \bowtie m_n(r) \bowtie 1$ from part (i) of the corollary. Thus $c_{t_n} \bowtie 1.$ Because total configurations are at least as numerous as root configurations, $m_n \leq c_{t_n}$, and then the growth of $m_n$ has exponential order at most that of $c_{t_n}$, so that $k_m \leq 1.$ Clearly, however, we cannot have $k_m<1$, because $m_n \geq 1$ for $n\geq 2$ and $k_m < 1$ would imply that the sequence $m_n$ decreases below 1 with increasing $n$. Thus, $k_m=1$.

For the sequence $M_n$, let $M_n \bowtie k_M^n$ be the exponential growth of the sequence $M_n$. This sequence has exponential order $k_M$. Suppose $(t_n)$ is any sequence of trees with $|t_n|=n$ such that $c_{t_n} = M_n$; that is, $t_n$ has the largest total number of configurations among trees of size $n$. From (\ref{doct}), $M_n \bowtie c_{t_n}(r)$, where the latter sequence has order smaller than or equal to $k_0$ because by definition  $M_n(r) \geq c_{t_n}(r)$ for all $n$, and $M_n(r) \bowtie k_0^n$ from part (i) of the corollary. Thus, $k_M \leq k_0.$ At the same time, for all $n$ we have $M_n \geq M_n(r)$, as the largest total number of configurations is larger than the largest number of root configurations. Thus, $k_M \geq k_0.$ It follows that $k = k_0$.


\subsection*{Appendix 4. Proof of (\ref{equaf})}

\begin{figure}
\begin{center}
\includegraphics*[scale=0.86,trim=0 0 0 0]{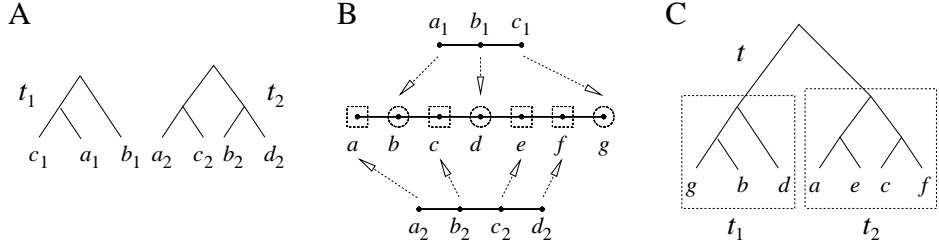}
\end{center}
\vspace{-.7cm}
\caption{{\small
Composition of two trees $t_1$ and $t_2$ of sizes $n_1=3$ and $n_2=4$ to obtain a tree $t$ of size $n=n_1+n_2=7$. {\bf (A)} Trees $t_1$ and $t_2$, with leaves labeled by $\{a_1,b_1,c_1 \}$ and $\{a_2,b_2,c_2,d_2  \}$. As in Section~\ref{filo}, we impose without loss of generality a linear order $a \prec b \prec c \prec \ldots $ for the leaves of a tree; here, we have $a_1 \prec b_1 \prec c_1$  and $a_2 \prec b_2 \prec c_2 \prec d_2$. {\bf (B)} Relabeling of trees $t_1$ and $t_2$. After relabeling, $t_1$ and $t_2$ have leaves labeled in the set $ a \prec b \prec c \prec d \prec e \prec f \prec g  $ of size $n=n_1+n_2$. For the relabeling procedure, we choose (dotted circles) $n_1$ elements among the $n$ possible new labels $\{a,b,c,d,e,f,g \}$. There are exactly ${{n}\choose{n_1}}$ different choices. The chosen elements relabel $t_1$, whereas the elements not selected (dotted squares) relabel $t_2$. With respect to the order $\prec$, the $i$th label of $t_1$ is assigned the label determined by the $i$th circle. Similarly, the $i$th label of $t_2$ is assigned the label determined by the $i$th square. {\bf (C)} After relabeling $t_1$ and $t_2$, the new tree $t$ is obtained by appending $t_1$ and $t_2$ to a shared root node. Starting with trees $t_1$ and $t_2$ in (A), the same procedure can generate ${{n}\choose{n_1}}$ different trees $t$, one for each possible choice of the $n_1$ elements (dotted circles) among the $n$ new labels. The only exception is when $t_1=t_2$, in which case the ${{n}\choose{n_1}}$ relabelings generate each tree exactly twice.}} \label{recu}
\end{figure}

The proof follows from the tree decomposition procedure that is illustrated in Fig.~\ref{recu}. According to this procedure, each tree $t$ of size $n$ is either the 1-taxon tree $t = \bullet$, or it can be created in a unique way by relabeling and appending to a shared root node two smaller trees $t_1$ and $t_2$ that become the root subtrees of $t$.
From Proposition~\ref{prop1}, the number $c_t(r)$ of root configurations of $t$ can be computed in this case as the product $[c_{t_1}(r)+1][c_{t_2}(r)+1]$. Summing over all possible trees $t$, the tree decomposition described in Fig.~\ref{recu} translates into the following decomposition for the generating function $F(z)$:
\begin{eqnarray}\nonumber
F(z) &=& \sum_{t \in T} \frac{c_t(r)}{|t|!} z^{|t|} = \frac{c_{\bullet}(r)}{1} z + \sum_{t: |t| > 1 } \frac{c_t(r)}{|t|!} z^{|t|} \\\label{pinca}
&=& \frac{c_{\bullet}(r)}{1} z + \frac{1}{2} \left[ \sum_{(t_1,t_2) \in \, T \times T} \frac{[c_{t_1}(r)+1][c_{t_2}(r)+1] z^{|t_1|+|t_2|}}{(|t_1|+|t_2|)!} {{|t_1|+|t_2|}\choose{|t_1|}} \right].
\end{eqnarray}
The first equality is the definition of $F(z)$. In the second equality, the set of trees over which the sum is evaluated is partitioned into two parts, the 1-taxon tree $t=\bullet$ and the trees of size larger than $1$. In the third equality, the set of trees $t$ with $|t|>1$ is realized taking all possible pairs of trees $(t_1,t_2) \in T \times T$, and applying to each pair the procedure in Fig.~\ref{recu}, considering all ${{|t_1|+|t_2|}\choose{|t_1|}}$ possible relabelings of $t_1$ and $t_2$. The quantity $c_t(r)$ in the sum $\sum_{t: |t| > 1 } \frac{c_t(r)}{|t|!} z^{|t|}$ is replaced by the product $c_t(r)=[c_{t_1}(r)+1][c_{t_2}(r)+1]$ and the term $|t|$ is replaced by $|t_1|+|t_2|$. Note the factor $1/2$ that appears in (\ref{pinca}) before the summation. This factor takes into account the fact that for each pair $(t_1,t_2) \in T \times T$ with $t_1\neq t_2$, there exists a symmetric pair $(t_2,t_1)$. Symmetric pairs generate exactly the same trees according to the procedure in Fig.~\ref{recu}, and multiplying by $1/2$ is required to avoid double counting. When $t_1=t_2$, the factor $1/2$ is still required because only half of the ${{|t_1|+|t_2|}\choose{|t_1|}}$ relabelings of $t_1$ and $t_2$ (Fig.~\ref{recu}B) create non-isomorphic trees when $t_1$ and $t_2$ are appended to a shared root node. Finally, observe that the number $c_{\bullet}(r)$ of root configurations in the 1-taxon tree is $0$.

From (\ref{pinca}) and the definitions of $F(z)$ and $T(z)$ in (\ref{genio}) and (\ref{eqFz}), algebraic manipulations yield
\begin{eqnarray}\nonumber
F(z) &=&  \frac{1}{2} \sum_{t_1 \in T} \sum_{t_2 \in T} \frac{[c_{t_1}(r)+1][c_{t_2}(r)+1] z^{|t_1|+|t_2|}}{(|t_1|+|t_2|)!} {{|t_1|+|t_2|}\choose{|t_1|}} \\\nonumber
&=& \frac{1}{2} \left(\sum_{t_1 \in T} \frac{c_{t_1}(r)}{|t_1|!}z^{|t_1|} + \sum_{t_1 \in T} \frac{z^{|t_1|}}{|t_1|!} \right)  \left(\sum_{t_2 \in T} \frac{c_{t_2}(r)}{|t_2|!}z^{|t_2|} + \sum_{t_2 \in T} \frac{z^{|t_2|}}{|t_2|!} \right)  \\\nonumber
&=& \frac{1}{2} [F(z) + T(z)]^2.
\end{eqnarray}


\subsection*{Appendix 5. Proof of (\ref{cappuccino})}
The proof follows the case of (\ref{equaf}). For $|t|>1$, the number $c_t(r)^2$ can be obtained as the product $(c_{t_1}(r)+1)^2(c_{t_2}(r)+1)^2$, where $t_1$ and $t_2$ are the root subtrees of $t$. The tree decomposition described in Fig.~\ref{recu} yields
\begin{eqnarray}\nonumber
G(z) &=& \sum_{t \in T} \frac{c_t(r)^2 }{|t|!} z^t = \frac{c_{\bullet}(r)^2}{1} z + \frac{1}{2} \left[ \sum_{(t_1,t_2) \in \, T \times T} \frac{[c_{t_1}(r)+1]^2[c_{t_2}(r)+1]^2 z^{|t_1|+|t_2|}}{(|t_1|+|t_2|)!} {{|t_1|+|t_2|}\choose{|t_1|}} \right] \\\nonumber
&=& \frac{1}{2}
\left( \sum_{t_1 \in T} \frac{c_{t_1}(r)^2z^{|t_1|}}{|t_1|!}
   + 2 \sum_{t_1 \in T} \frac{c_{t_1}(r)z^{|t_1|}}{|t_1|!}
   +   \sum_{t_1 \in T} \frac{z^{|t_1|}}{|t_1|!} \right)
\left( \sum_{t_2 \in T} \frac{c_{t_2}(r)^2z^{|t_2|}}{|t_2|!}
   + 2 \sum_{t_2 \in T} \frac{c_{t_2}(r)z^{|t_2|}}{|t_2|!}
   +   \sum_{t_2 \in T} \frac{z^{|t_2|}}{|t_2|!} \right)
\\\nonumber
&=& \frac{1}{2} [G(z) + 2F(z) + T(z)]^2.
\end{eqnarray}

\noindent
{\footnotesize
{\bf Acknowledgments.}
We thank Elizabeth Allman, James Degnan, and John Rhodes for discussions and NIH grant R01 GM117590 for financial support.}

\bibliographystyle{chicago}
\bibliography{map3}

\end{document}